\documentstyle[aps,prl,multicol,epsf,psfig,boxedminipage,citesort,graphicx]{revtex}  

\newcommand{\be}{\begin{equation}}
\newcommand{\bel}[1]{\begin{equation}\label{#1}}
\newcommand{\ee}{\end{equation}}
\newcommand{\bea}{\begin{eqnarray}}
\newcommand{\ba}{\begin{array}}
\newcommand{\eea}{\end{eqnarray}}
\newcommand{\ea}{\end{array}}

\newcommand{\DIR}{.}

\begin{document}

\twocolumn[\hsize\textwidth\columnwidth\hsize\csname@twocolumnfalse%
\endcsname

\title{ Human behavior as origin of traffic phases}

\author{Wolfgang Knospe$^1$, Ludger Santen$^2$, Andreas Schadschneider$^3$ 
and Michael Schreckenberg$^1$}

\address{$^1$ Theoretische Physik FB 10,
          Gerhard-Mercator-Universit\"at Duisburg,
                  Lotharstr. 1, D-47048 Duisburg, Germany}

\address{$^2$ CNRS-Laboratoire de Physique Statistique, Ecole Normale
Sup{\'{e}}rieure, 24, rue Lhomond, F-75231 Paris Cedex 05, France}

\address{$^3$ Institut f\"ur Theoretische Physik,
                 Universit\"at zu K\"oln,
                 Z\"ulpicher Str. 77, D-50937 K\"oln, Germany }

\date{\today}

\maketitle

\begin{abstract}
It is shown that the desire for smooth and comfortable driving is directly
responsible for the occurrence of complex spatio-temporal structures
(``synchronized traffic'') in highway traffic. 
This desire goes beyond the avoidance of accidents which so far has been the
main focus of microscopic modeling and which is mainly responsible
for the other two phases observed empirically, free flow and
wide moving jams.
These features have been incorporated into a microscopic model based on 
stochastic cellular automata and the results of computer
simulations are compared with empirical data.
The simple structure of the model allows for very fast implementations
of realistic  networks. The level of agreement with the empirical findings 
opens new perspectives for reliable traffic forecasts.  
\end{abstract}

\bigskip
]

The empirical observation of highway traffic has shown the existence of
very complex spatio-temporal structures
\cite{empiric}. 
Therefore it was questioned if it is possible to find a model 
which  can reproduce the various observed phenomena and 
which is able to relate the origin of these phenomena to the human
driving behavior. Compared to the analysis of  
a conventional liquid the difficulties in modeling traffic are already in 
describing the ``microscopic'' interactions between 
the vehicles. These do not follow the fundamental laws of nature but are 
rather complex decisions taken individually by each driver. 
Therefore any successful microscopic description has to extract 
the essentials of the human behavior in different traffic situations, 
which is clearly a difficult task. Most present modeling approaches
concentrate on  
the fact that drivers want to avoid accidents. This has been implemented 
by adjusting the velocity
according to differences in speed and/or the distances to the other
cars. Traffic models based on this interaction only have been
able to reproduce various observed phenomena but fail to give a
complete description of the empirical
results\cite{review}.
Much progress has been made in recent years in understanding 
topological effects in highway traffic~\cite{treiber00}. It became
clear that e.g.~a change of the activity of an on-ramp may induce a
phase transition 
on the highway. The observed phase transitions are between  
stable traffic states, i.e. the generated traffic states dissolve only 
if the external conditions change again. The robustness of the empirical
observations give strong evidence that traffic states itself are 
a consequence of the human driving behavior rather than a response 
to different topological situations.

Below we will show that it is 
possible to overcome the problems in modeling traffic if one
takes into account that 
people like to have a comfortable journey, i.e., they try to avoid strong
accelerations and abrupt braking. This second essential demand of drivers 
has been incorporated in a recent traffic model\cite{mod_jpa}, 
which introduces ``brake-lights'' for a timely adjustment of the velocity when
approaching slow upstream traffic and ``anticipation'' by estimating
the velocity of the leading vehicle. 
This approach goes far beyond the consideration of velocity
differences since acceleration changes become visible and allows for an
event driven anticipation of velocity reductions.
These two features lead to a
stabilization of the flow in dense traffic which is crucial to
overcome the difficulties in describing the empirically observed
phases and their transitions\cite{empiric}.
Like any other theoretical approach traffic models have to be
compared with empirical results. Here one has the difficulty that one
cannot perform experiments but is restricted to pure observation of a
given situation. Nevertheless much progress has been made in recent years
by analyzing large data sets in different environments. So it is now
widely believed that at least three traffic states exist, i.e., {\em
(i)} free-flow 
{\em (ii)} synchronized traffic and {\em (iii)} wide moving
jams~\cite{atleast}.  
The
characteristics of 
free-flow~\cite{free} and wide moving jams~\cite{sag}
are intuitively
clear. For a long time it was believed that these states are the only
stable traffic states. This commonly accepted picture was enhanced  by
establishing a second stable congested state, i.e., synchronized
traffic.   
Synchronized traffic~\cite{empiric}, which is typically observed at
on- and off-ramps,  is characterized by a large 
variance in flow and density measurements and a velocity which is
significantly lower than in free-flow traffic. 
The origin of the notation ``synchronized traffic'' is the fact that 
the time-series of measurements on different lanes are highly correlated. 
But more important 
is the apparent absence of a functional flow-density form, i.e., the 
measurements of the flow, density and velocity of the traffic are 
distributed over a wide area~\cite{kerner_synchro}. 
This observation has been confirmed
quantitatively by means of vanishing cross-correlations between these
two quantities. 
Synchronized traffic and wide moving jams differ also in their behavior 
at bottlenecks. If synchronized flow is generated at a bottleneck its 
downstream front is pinned at the bottleneck. In contrast the downstream 
front of wide moving jams propagates with constant velocity in upstream 
direction. Their velocity does not depend on the traffic state they cross and
is even unchanged if they pass a bottleneck.
The final characteristic feature are the phase
transitions between the different states~\cite{empiric}. In  
general they are of first order~\cite{fot}, e.g., the phase transition
free-flow 
to synchronized traffic is characterized by a discontinuous change of 
the velocity. 

\begin{figure} 
\begin{center}
\includegraphics[width=0.75\linewidth]{\DIR/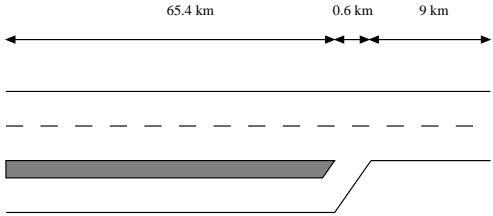}
\end{center}
\caption{Schematic plot of the highway section modeled throughout
this work. The space has been discretized such that each lattice
site corresponds to 1.5 m in reality. The total length of the highway
is 75~km or 50,000 cells per lane. The merging zone of the on-ramp is of
length 600 m, i.e., 400 cells. Lane-changing rules as presented in
[13,14] have been used. The  results,
 however, do not depend on the
details of the applied traffic rules. We simulate fluid traffic on the
on-ramp with an average velocity of $\sim 80 km/h$. In 
addition the incoming cars accept smaller gaps for 
lane-changes [15]. } 
\label{fig:segment}
\end{figure}

Already the correct reproduction of synchronized traffic on a
quantitative level is a difficult 
task\cite{synchro}, but the most puzzling
point for any model is to reproduce stable synchronized traffic in
coexistence with the other two traffic states\cite{empiric}. 
The discussed model is able to pass this most sensitive test. 

The model~\cite{mod_jpa} is based on the cellular automaton model
of Nagel and Schreckenberg~\cite{nasch} and  
incorporates the desire for smooth and comfortable driving by 
anticipating the velocity of the leading vehicle and the introduction
of ''brake lights''.
 For comparison with the empirical data we
 simulate a two-lane segment with an on-ramp (fig.~\ref{fig:segment}). 
The additional input of cars 
triggers a dense
traffic region behind the on-ramp which can be identified as
synchronized traffic. Our simulations clarify how the synchronized
state is related to the human factors in driving.

{\em (i) Velocity anticipation:} At the 
on-ramp the anticipation of the leaders velocity avoids abrupt braking
of the traffic behind and therefore reduces the probability to form
jams.

{\em (ii) Retarded acceleration:} Comfortable driving also implies
that cars do not accelerate immediately in case of a larger gap ahead
if they observe slow downstream traffic. On the one hand, this leads
in some sense to a sub optimal gap usage, because the velocity is
smaller than the headway allows. On the other hand, larger
gaps in a dense region reduce the car-car interactions what may cut a
chain of braking overreactions which is responsible for the formation
of jams. These overreactions are a direct consequence of the delayed
human behavior in adapting the velocity to the headway which can
lead to an avalanche like amplification of the velocity fluctuations
upstream and finally to the formation of jams.

{\em (iii) Timely braking:} Finally timely braking
suppresses another mechanism of jam formation: 
When the velocity adjustment is only based on the distance to the
next car ahead, jams often emerge in the layer between free-flow and 
synchronized traffic. 
In these models the jam formation arises from cars approaching
a slow-moving cluster with high speed which leads to a compactified
region. In contrast, our approach avoids this 
artificial mechanism to form a jam, the drivers adjust their speed to
the vehicles ahead.

\begin{figure}[h]
\begin{center}
\includegraphics[width=0.8\linewidth]{\DIR/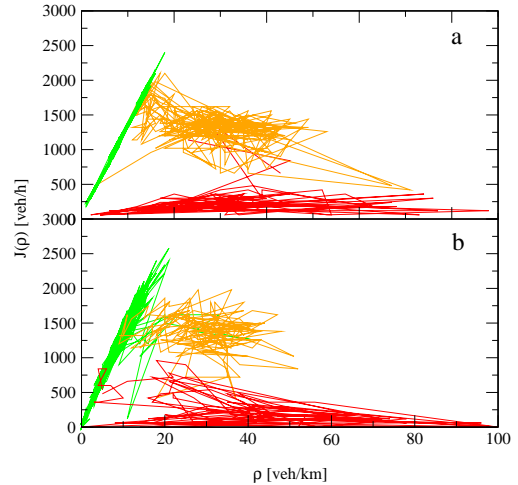}
\caption{Comparison between empirical results (b) for the flow-density
relation (fundamental diagram), and time-traced simulation
results (a). On the abscissa the density (the number of vehicles per
kilometer) and on the ordinate the flow (number of cars passing the
detector extrapolated to one hour) is shown. Each data point corresponds to an 
average over an one-minute interval. Consecutive measurements are
connected by lines. Part (a) shows that we recover the three empirically
observed phases of highway traffic: free-flow (green), synchronized 
traffic (orange) and wide jams (red). 
The empirical data are one-minute
averages of detector data from the German freeway A40 near Moers
junction 
at 2000-12-12 (synchronized state) and near Bochum-Werne junction at
2001-02-14 (wide moving jam of about 2 hours duration).   
The density is given by the occupancy which gives the
percentage of the measurement time a detector is covered by cars. This
has the advantage that the occupancy can directly be measured by an
inductive loop and the simulation data can be related to the empirical
data.} 
\label{fig:abtrans}
\end{center}
\end{figure}

At this point we stress the fact that 
the observed phenomena are a
consequence of the individual behavior of the drivers 
(See~\cite{mod_jpa,parameter} for the calibration of the model).
None of the microscopic model parameters has been changed throughout the 
simulations in order to optimize the accordance with the empirics.
Already the single lane model on a periodic street without 
bottlenecks~\cite{mod_jpa} shows the 
existence of synchronized traffic and wide moving jams.
In this
simulation the boundary conditions are used only to 
induce the transitions to these traffic states.
By contrast, the excellent agreement with the empirics was obtained
simply by applying  
the correct inflow at the upstream end and the on-ramp of the highway 
section. This side-steps another important question in traffic dynamics, 
i.e., the origin of phase transitions. Our simulation results support the 
view that the transitions are mostly induced by obstacles rather than 
by a spontaneous breakdown of the traffic stream.

The simulation protocol emulates a few hours of highway traffic
including the realistic variations of the number of cars which are fed
into the system. A busy on-ramp in combination with high flow
generates in general synchronized flow on the 
highway segment. For the sake of simplicity we used only one type of
cars in the simulations which leads to a smaller variance of the data
points in the free flow regime compared with the empirical data.

\begin{figure}[h]
\begin{center}
\includegraphics[width=\linewidth]{\DIR/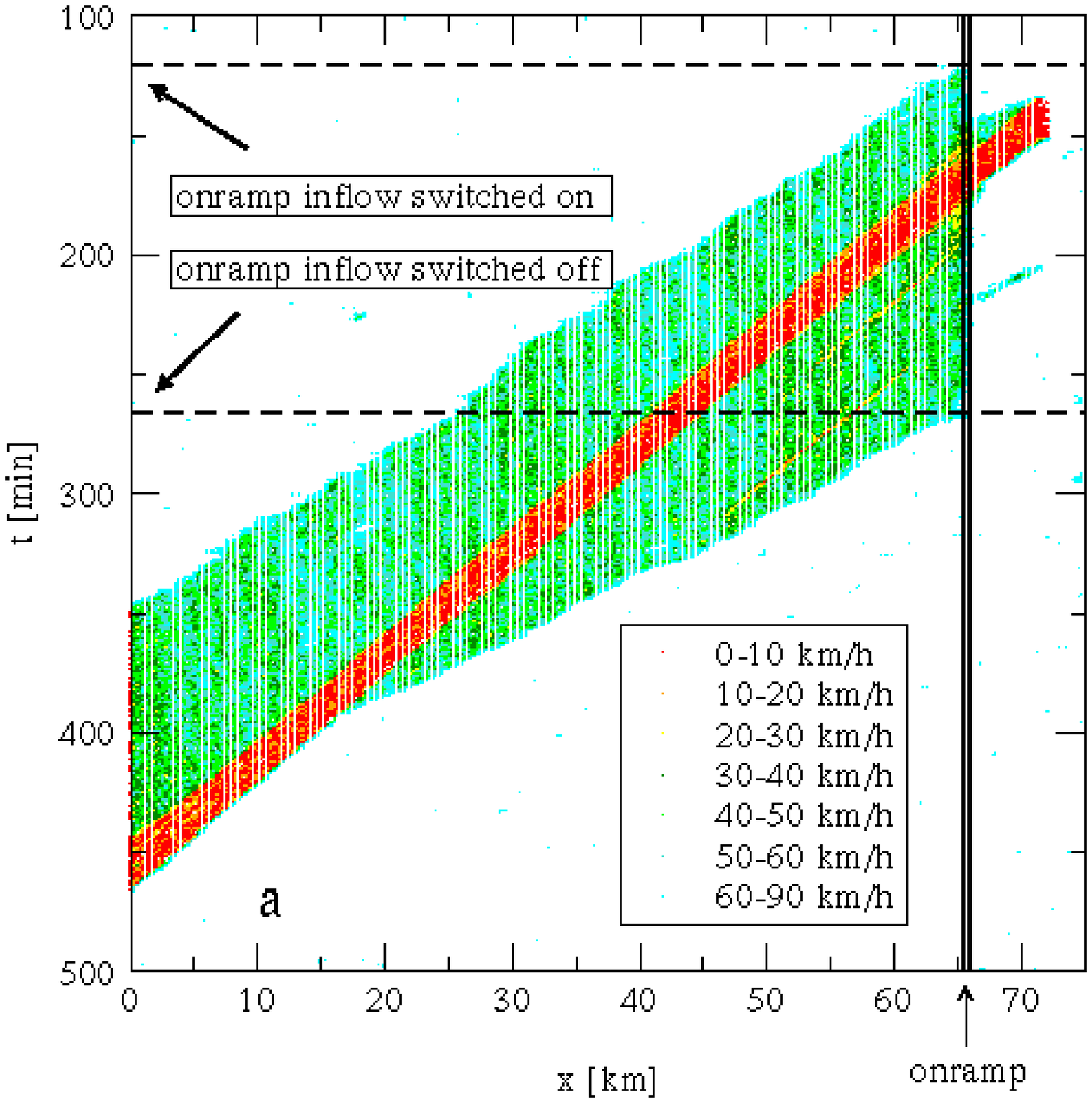}
\includegraphics[width=\linewidth]{\DIR/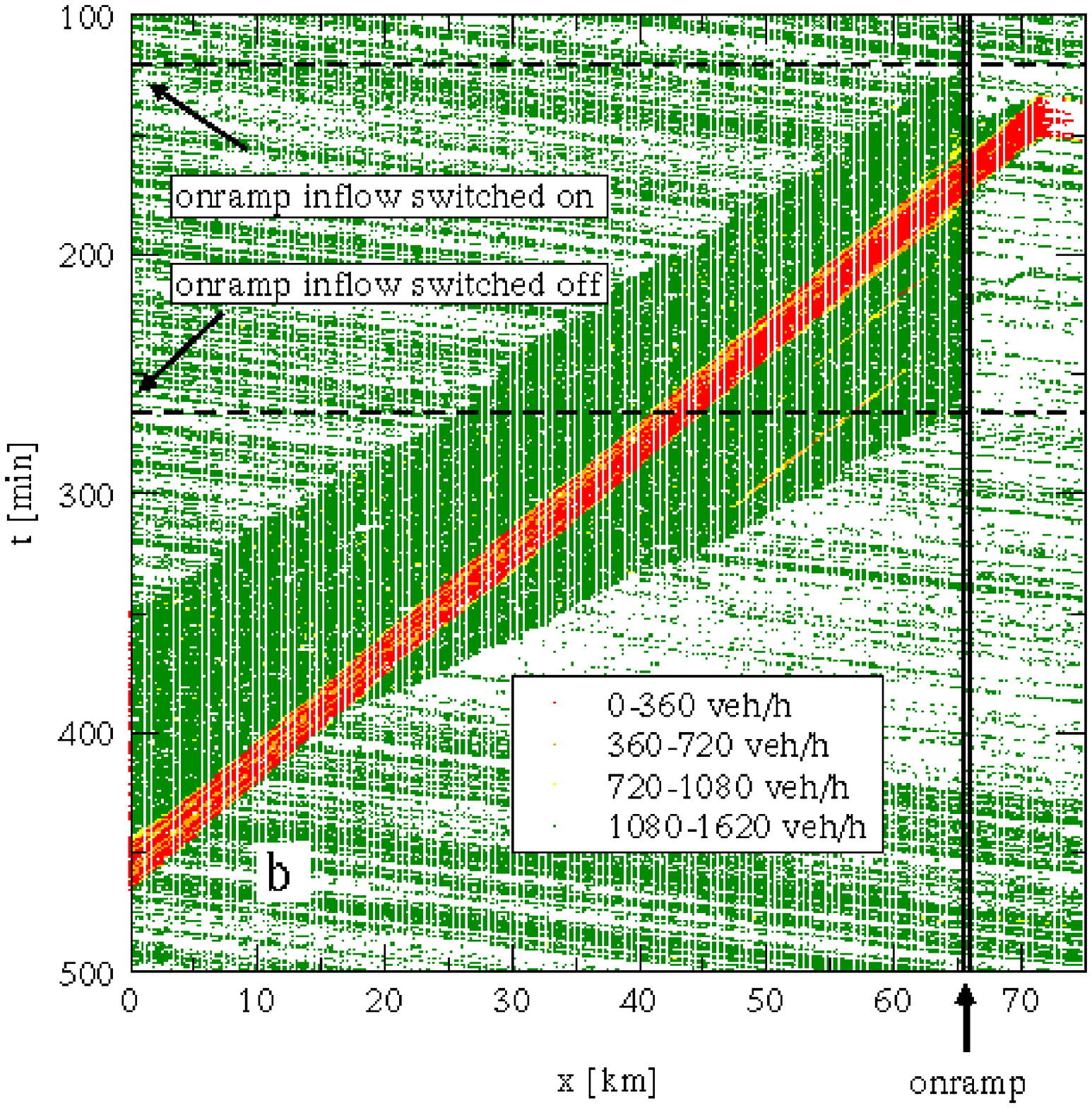}
\caption{Coexistence of wide moving jams and synchronized states. 
Space/time  evolution of the velocity (a) and of the flow (b). 
The figures show how a
traveling jam wave crosses a region of synchronized traffic which is
pinned at the on-ramp.  Downstream the on-ramp a jam has been generated which
moves in upstream direction and passes the segment with 
free flow and synchronized states. One clearly observes that the
synchronized state is recovered directly after the jam has passed the
on-ramp.}
\label{fig:st}
\end{center}
\end{figure}

The simulations show that we
can recover the empirical results for the fundamental diagram
quantitatively (see Fig.\ref{fig:abtrans}). The agreement is not only
for the average values but also concern the statistical properties 
of the results. This is mandatory for traffic forecasting, e.g., in 
order to calculate upper limits of individual travel times as well. 
But, as mentioned above, also the stability of the synchronized
traffic state
is described correctly. In order to verify this a jam was generated
by an obstacle at the downstream end of the highway section.
Figure~\ref{fig:st} illustrates how the jam wave travels through
the free flow region with constant velocity and also passes the
section were the 
synchronized traffic is localized. This shows that we can superpose
the different traffic states as observed empirically.
The fact that the jam propagates with a constant velocity and
passes the free-flow, on-ramp and synchronized regions without
being disturbed is typical for wide moving jams.

Our simulation results have far reaching consequences for the theory
of traffic flow as well as for practical applications. 
From a theoretical point of view, it is shown that the desire for
smooth and comfortable driving is 
the origin of synchronized traffic and wide moving jams and is
responsible for the stability of the
different traffic phases. This stability allows
for the application of phenomenological approaches. 
In particular the motion and formation of jam waves, which is most
interesting for any traffic forecast, should be predictable within
these approaches with high accuracy (see \cite{poppy}
for approaches of this kind).  
  
From a practical point of view, 
our simulation results open the door
the desire for smooth and comfortable driving allows for more
realistic simulations and opens the door
for a forecast of highway traffic which should outperform knowledge
based approaches. 
Models of this type have
already been used for fast microsimulations of large traffic
networks.
Here the simple structure of the model, i.e., its discreteness
and the local rules, is of great importance. 
   
Summarizing, we have shown that a rather simple cellular automaton model
is able to reproduce the empirically observed phases of traffic flow even
quantitatively. The model parameters can be related directly to the human
behaviour, especially to the desire for smooth and comfortable driving.
This desire turns out to be responsible for the occurance of complex
spatio-temporal structures, like the so-called synchronized traffic
phase.

{\bf Acknowledgments}: 
We thank D.~Helbing and B.S. Kerner for useful discussions. 
The authors are grateful to the Landesbetrieb Stra\ss enbau NRW for
the data support and to the 
Ministry of Economics and Midsize Businesses, Technology and Transport
for the financial support.
L.~S.~acknowledges support from the Deutsche Forschungsgemeinschaft
under Grant No. SA864/1-2.  

\bibliographystyle{unsrt}

\begin{thebibliography}{99}
\bibitem{atleast} More phases might be distinguished in the
presence of external 
disturbances~\cite{treiber00}, e.~g., ramps.
\bibitem{free} Free flow traffic is characterized by a high average
velocity of the vehicles, e.g., close to the eventually applied speed
limit.
\bibitem{sag} Wide moving traffic jams are upstream moving
structures consisting of two fronts separated by a region of
negligible velocity and flow. They can be characterized by the
upstream propagation velocity, their density and the
outflow from the jam. These parameters of the jam are quite 
robust: Empirically it has been observed that two jams can move in parallel 
through the system which implies that their characteristic parameters agree.
\bibitem{parameter} The calibration of the microscopic driving
behavior resulted in the following  model
parameters: $v_{\text{max}} = 18$, $p_d = 0.1$, $p_b = 0.95$, $p_0 = 0.5$, $h
= 8$, $gap_{\text{security}} = 3$. The cell length is $1.5~m$, a car has a
length of $5$ cells (see \cite{mod_jpa} for a detailed discussion of
the rules and model parameters).
\bibitem{twolane_our_approach} We modified the symmetric lane changing rule set
of~\cite{twolane_rickert} by including anticipation effects for the
calculation of an effective gap to the predecessor on the destination
lane and weakened the security criterion for the distance to the
successor on the destination lane to the velocity of the successor.
\bibitem{twolane_onramp} Lane changing rules
of~\cite{twolane_rickert} are applied. A car is allowed to change
the lane if its has an effective gap of 5 cells to the predecessor and
a gap of 5 cells to the successor on the
destination lane. Anticipation of the velocity of the predecessor on
the destination lane increases the gap and leads to an effective gap.
Additionally, we dropped the incentive criterion.
\end{thebibliography}

\end{document}